\documentclass[journal]{IEEEtran}

\usepackage{cite}

\usepackage{graphicx}

\usepackage[usenames, dvipsnames]{color}

\begin{document}

\title{Physical Education and English Language Arts Based K-12 Engineering
Outreach in Software Defined Networking (Extended Version)
\thanks{This work was supported by the National Science Foundation through Grant $1716121$.}}

\author{Gamze~Ozogul,
Akhilesh~S.~Thyagaturu,~\IEEEmembership{Member,~IEEE,}
Martin~Reisslein,~\IEEEmembership{Fellow,~IEEE,}
and~Anna~Scaglione,~\IEEEmembership{Fellow,~IEEE}
\thanks{G.~Ozogul is with the School of Education,
Indiana University Bloomington (e-mail: gozogul@indiana.edu).}
\thanks{A. Thyagaturu is with the Programmable Solutions Group (PSG),
Intel Corporation, Chandler, AZ 85226, USA (e-mail: akhilesh.s.thyagaturu@intel.com).}
\thanks{M.~Reisslein and A.~Scaglione are with the
School of Electrical, Computer, and Energy Engineering,
Arizona State University (ASU), Tempe, AZ 85287, USA (e-mail: \{reisslein, ascaglio\}@asu.edu).}}

\markboth{Ozogul \MakeLowercase{\textit{et al.}}:
PE and ELA Based K-$12$ Engineering Outreach in SDN}%
{Ozogul \MakeLowercase{\textit{et al.}}:
PE and ELA Based K-$12$ Engineering Outreach in SDN}

\maketitle

\begin{abstract}
K-12 engineering outreach has typically focused on elementary
electrical and mechanical engineering or robot experiments
integrated in science or math classes. In contrast, we propose a
novel outreach program focusing on communication network principles
that enable the ubiquitous web and smart-phone applications. We
design outreach activities that illustrate the communication network
principles through activities and team competitions in physical
education (PE) as well as story writing and cartooning in English
Language Arts (ELA) classes.  The PE activities cover the principles
of store-and-forward packet switching, Hypertext Transfer Protocol
(HTTP) web page download, connection establishment in cellular
wireless networks, as well as packet routing in Software-Defined
Networking (SDN).  The proposed outreach program has been
formatively evaluated by K-12 teachers.  A survey for
the evaluation of the impact of the outreach program on the student
perceptions, specifically, the students' interest, self-efficacy,
utility, and negative stereotype perceptions towards communication
network engineering, is also presented.
\end{abstract}


\IEEEpeerreviewmaketitle

\section{Introduction}

\subsection{Existing Approaches: Mainly Basic Engineering in Math and Science Classes}
Early engagement of K-$12$ students with engineering is widely
considered to be critical for creating interest in engineering
programs of study at colleges and
universities~\cite{ada2011mul,bro2008adv,bug2017eng,car2012eng,dua2018bas,gar2020joy,gre2017des,hol2019inf,inc2009est,kel2019epi,moo2015ngs,ozo2016k,sim2006mat,tab2018dev}.
Engaging pre-college students with fun-filled, yet educational
activities has shown benefits in increasing student interest in
engineering~\cite{cao2019dis,ham2019lig,inn2012ari,mel2019eng,moo2014fra,mos2007k,sun2011eng,sun2015eng,sun2019ste,ozo2019sch}.

Most engineering outreach activities to date have focused on
elementary electrical and mechanical engineering topics, such as light
bulb circuits and bridge
building~\cite{jef2004und,joh2013ped,ozo2013inv,rei2010pre} or
robotics~\cite{anw2019sys,cej2006kin,riv2017eng}.  Moreover, engineering outreach
activities conducted in school classrooms have so far mainly been
integrated into science and mathematics
classes~\cite{hun2006opp,mos2011sup}.

\subsection{Novel Approach: Communication Networks in
  Physical Education and English Language Arts Classes}

Only a few engineering outreach activities have presented elements of
modern wireless communication and networking (see
e.g.,~\cite{aba2013con,bak2010app,lim2010app,spa2016dev}), Internet of
Things concepts (see e.g.,~\cite{kin2019sop,mor2018iot,sie2019mic}),
and general computing device concepts (see e.g.,~\cite{ruc2018els}) to
K-12 students.  Given the widespread use of wireless communications
services by youth today, it appears critical to engage $K-12$ students
with the basic principles of modern wireless communication and
networking systems.

More specifically, communication networks are the underlying enabling
technology for a wide variety of information technology applications
that K-$12$ students interact with every day. For instance, the
ubiquitous smart phones and the wide range of services that they
provide are enabled by underlying communication network architectures
and protocols. The increasing adoption of smart phones and new
wireless communication standards~\cite{nav2020sur} enable high-quality
multimedia experiences with high responsiveness. The
underlying communication network architectures and protocols require
careful engineering to achieve these low-latency high-bitrate
services. One important component towards attaining and further
enhancing these services is software-defined networking (SDN), which
allows for the fine-grained control and optimization of the resource
usage in communication networks~\cite{kel2019ada}.  A related strategy
is to cache frequently requested content close to the end users and to
conduct service computing, e.g., for queries and security
applications, close to the users in so-called multi-access edge
computing (MEC)
infrastructures~\cite{fer2019tow,jun2020rec,meh2019dev,she2020sec,xia2019red}.
SDN provides a framework to control these communication, caching, and
computing resources in a wide variety of communication contexts,
including wireless networks and so-called backhaul networks that
connect wireless networks to the wired Internet at
large~\cite{sha2018lay,thy2016sdn,thy2018r}.

Following the emerging educational paradigm of integrated
project-based learning activities across seemingly disparate
subjects~\cite{kle2005int,lou2011imp}, we integrate the communication
network engineering outreach activities in Physical Education (PE) and
English Language Arts (ELA) instruction.  General principles of
integrating science, technology, engineering, and mathematics (STEM)
or science, technology, engineering, arts, and mathematics (STEAM)
activities in PE instruction have recently attracted significant
interest, see
e.g.,~\cite{cho2018eff,erw2017ful,coe2020ste,har2020man,mar2016fit,waj2019ste}.
Also, the relevance of biomechanics STEM concepts for youth athletes
have recently been explored~\cite{dra2020bio}.  Moreover, the
integration of project based learning into PE instructions has been
considered in~\cite{tre2018mak}.  Engineering outreach with the Xbox
$360$~Kinect system has been explored in~\cite{bla2013inn}, while a
trebuchet (catapult) competition that seeks to connect engineering
with a sports activity has been described in~\cite{sla2008mak}.
A scavenger hunt with wireless sensor network components has been
studied in~\cite{wah2011tec}.

A concept for integrating engineering design into elementary school
English literacy has been examined
in~\cite{mcc2016sta,mon2019nov,por2015nov,por2018nov}.

\subsection{Contributions and Structure of this Article}
We design communication network engineering outreach activities that
are primarily targeted for middle schools (grades $6$--$8$); however, high
elementary school grades, e.g., grades $4$ and $5$, as well as low high
school grades, e.g., grades $9$ and $10$ may also enjoy and benefit from
the designed outreach activities.  The designed student activities
specifically fit the middle school age group as well as adjacent
elementary and high school grades. We have tailored the level of
technical detail as well as the presentation of background knowledge
to prepare the students to conduct meaningful network engineering
activities.

The designed outreach activities are interdisciplinary, bringing the
dynamics of communication networks to life in activities in
conventional physical education (PE) and English Language Arts (ELA)
middle school classes. The PE activities are described in
Section~\ref{peact:sec} while the ELA activities are described in
Section~\ref{elac:sec}. The designed PE activities cover crucial
communication networking principles ranging from store-and-forward
packet switching and web page downloading to connection establishment
in wireless cellular networks and routing in SDN vs.~Internet Protocol
(IP) networks. The activities are designed to allow teams of students,
e.g., two halves of a PE class or different PE classes to compete
against each other in races that simulate the communication network
operations.  The designed ELA activities ask students to ``personify''
the communication network equipment and data packets and to write
stories or cartoons that exemplify communication network operations.

We have conducted formative evaluations of the designed outreach
activities with K-$12$ teachers as
described in Section~\ref{eval:sec}.  Towards rigorously evaluating the
impact of the designed outreach activities on student interest,
self-efficacy, stereotypes, and utility perceptions of the network
engineering field, we have adapted a validated evaluation survey.

\section{PE Activities to Model Communication Network Operation}
\label{peact:sec}
\subsection{Basic Modeling Principles}

\subsubsection{Brief Background on Communication Networks}
A communication network consists mainly of a network architecture that
is operated according to a prescribed set of networking protocols.
The network architecture consists of network nodes that are fixed
(stationary), e.g., routers or SDN switches in a metropolitan
area~\cite{alf2017mon,mai2004awg,yan2004met} or backbone network, or
mobile, e.g., wireless phones and laptops.  The network nodes are
interconnected by communication links with a prescribed topology
(connectivity pattern). A communication link is characterized by a
transmission bitrate in units of bit per second, indicating how many
bits can be transmitted per second onto the link, and by a propagation
speed in units of meters per second, indicating how fast the physical
signal carrying the bit information travels (propagates) along the
link.

A communication protocol gives rules for the operation of the
communication network. That is, the communication protocol prescribes
the sequences of messages that are exchanged by communication nodes as
well as the actions that nodes take upon the receipt of certain
messages.

The application that everyday users interact with, e.g., e-mail,
multi-player gaming, web browsing, and video streaming accomplish their
services to the human user through the exchange of discrete data
messages or continuous data traffic flows over the network of
communication nodes that makes up the
Internet~\cite{jul2015mea,rei1998hig,li2008con,li2008ene}.  Data
messages and continuous data traffic flows are typically carried via
discrete packets, e.g., an small text e-mail (message) can be carried
in one data packet, while a large image or video file that is attached
to an e-mail may require several packets to carry all the image or
video data~\cite{see2014vid,tan2013sur}. A data packet is
characterized by a packet size in units of bit or byte (whereby 8 bit
correspond to one byte).

\subsubsection{Representing Communication Network Components Through
  Student Actions}
\paragraph{Overview} The basic principle of our proposed outreach
activities is to have students physically move to model (represent)
the interactions of the components in a communication network. This
section describes general principles and options for this
representation of communication network operations through students
actions.  The following sections then illustrate these basic
principles in the context of specific examples of common
communication network operations.

\paragraph{Network Node}
\begin{figure}[t!]  \centering
\includegraphics[width=3.5in]{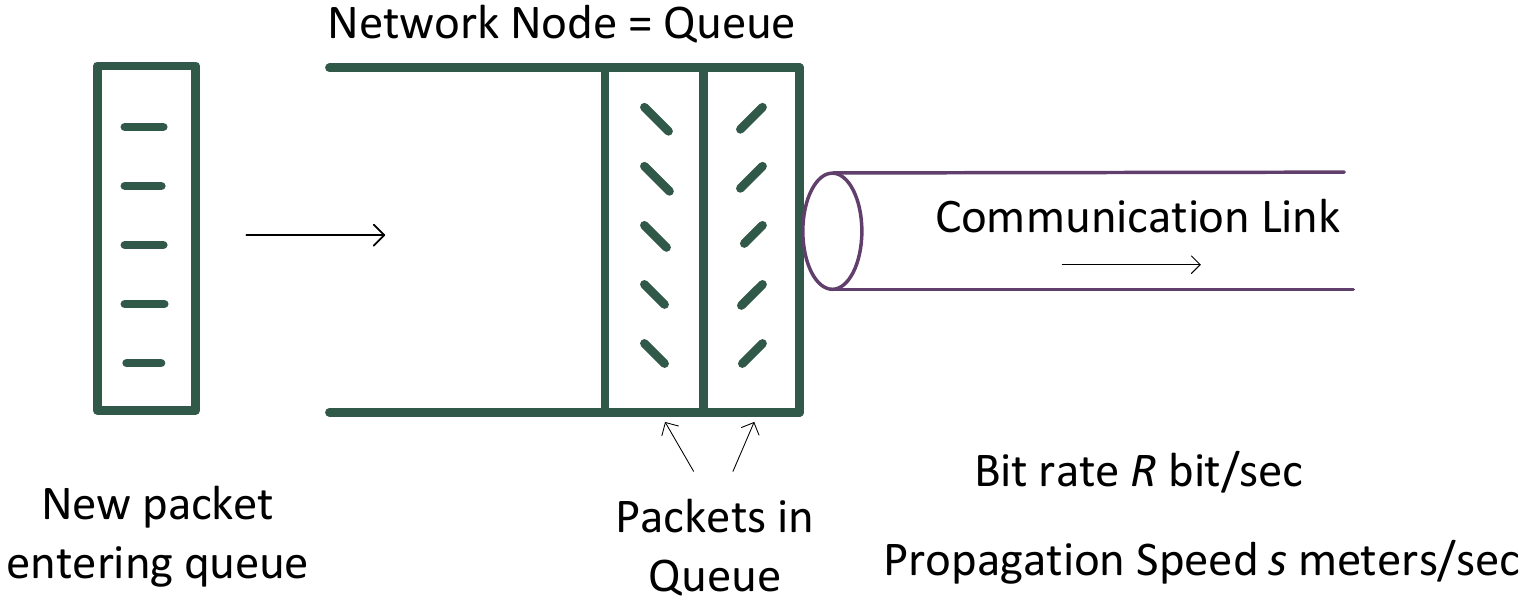}
\caption{Network node represented by a queue that holds several
  packets. The packets are transmitted over the attached communication
  link at bitrate $R$~bit/second; the bit signals travel with
  propagation speed $s$~meters/second over the link to the next network
  node or a destination. }
  \label{node:fig}
\end{figure}
Generally, a network node, e.g., router or switch, can be represented
by a student that is assigned to carry out the tasks of the node,
e.g., by directing the actions of other students that represent the
data traffic packets.  The modeling of stationary network nodes,
presents an opportunity for integrating students with physical
limitations, e.g., special needs students with mobility limitations or
students with limited mobility due to injuries, into the outreach
activities. Alternatively, a network node could be represented through
a ``queue'' symbol (see Fig.~\ref{node:fig}), i.e., a rectangle that
is open on a short side to let the bits or packets enter and long
enough to ``hold'' the maximum number of bits that occur in a data
transfer. The queue symbol can be marked on the ground, e.g., through
chalk or some other erasable marking.

\paragraph{Communication Link}
The communication link can be represented through a line that is drawn
on the ground to ``connect'' one node, in particular, the closed short
side of the queue symbol of a node to the open short side of the queue symbol
that represents the next node.  The characteristics of the link, i.e.,
the transmission bitrate and propagation speed can be enforced by a
student that represents the node, i.e., directs the actions of the
students representing the bits or packets. For instance, the ``node
student'' could let one ``bit student'' depart per second to model a
transmission bit rate of $1$~bit per second, i.e., $1$ student per
second. The ``node student'' could also instruct the ``bit student'' how
fast to propagate along the line representing the link, e.g., whether
to walk slowly, walk normally, walk briskly, jog, run, or sprint.
Alternatively, if no student is assigned to represent a communication
node, then the link characteristics could be written on a sign that is
posted by the closed short side of the queue symbol.

\paragraph{Data Message and Packet}
A data packet can be modeled by one student that represents the entire
packet. Alternatively, a given data packet of $P$~bits (with $P \geq
1$) can be represented by $P$ students that need to traverse the
network in a group. Possibly, multiple packets are needed to represent
an application layer message, e.g., a message could consist of
$M$~bits, with $M = k P$ for some positive integer $k$.

An enriched data packet model can utilize some artifact, such as a
bean bag, a piece of a jigsaw (tiling) puzzle, or a bucket of water,
to represent a data packet. The individual data packet representations
can then be assembled to give a message, e.g., by collecting a
prescribed number of bean bags, putting together a jigsaw puzzle, or
filling a tub with a prescribed amount of water.

\subsection{Modeling of Specific Communication Network Operations}
This section describes the modeling of specific communication network
operations through PE activities of students.  We proceed in a
didactical manner from elementary store-and-forward packet switching,
which is the basic operating principle of the Internet to a more
complex operating principle, namely HTML web page download.  We
furthermore describe a PE activity to model the connection
establishment in a wireless network.  Finally, we outline a PE
activity to model the interactions between the control plane and data
plane in Software-Defined Networking (SDN) in the context of packet
routing, which is the basis for a wide range of optimizations and
enhancements that make modern ubiquitous multimedia networking
possible.

\subsubsection{Store-and-Forward Packet Switching} \label{stoforps:sec}
\paragraph{Background}
Consider the transmission of a message of $M$~bits over a linear
sequence of networking nodes, specifically, from a sender (source)
\texttt{A} to a destination \texttt{B} via $N$ intermediate
packet-switching network nodes (routers or switches), i.e., via $N +
1$ interconnecting communication links. Suppose that each link has
bitrate $R$~bit/s and a propagation speed $s$ meters per second and a
length $l$ in meters.  Suppose that the message can be partitioned
into $M / P$ packets, where $P$ denotes the packet size in bits, and
$M/P$ is assumed to be a positive integer.

The basic principle of store-and-forward packet switching requires
that all $P$ bits are received at an intermediate node before the node
can forward the packet to the next network node. Thus, a packet incurs
a delay or $P/R$ seconds to be transmitted by the source node and a
propagation delay of $l/s$ seconds to reach the first intermediate
node. The first intermediate node needs to assemble all the $P$ bits
in the packet before the node can process the
packet~\cite{for2012com,kur2017com}. We assume that the processing
delay is negligible. In practice, there is a very short delay,
typically on the order of nanoseconds or less to determine the how to
forward the packet (in our linear network example, all packets
progress linearly through the linear sequence of network
nodes).

The first intermediate network node then transmits the packet onto the
link to the second intermediate network node (transmission delay
$P/R$) and the bits propagate along the link to the second
intermediate network node (propagation delay $l/s$).  This process
repeats for all $N$ intermediate network nodes, for a total delay of
$(N+1) \left[ \frac{P}{R} + \frac{l}{s} \right]$ from the time instant
of the start of the transmission at \texttt{A} to the complete
reception of the first packet at \texttt{B}.

\paragraph{Student Modeling}  \label{stumodps:sec}
This store-and-forward packet transmission process can be modeled
with students as follows. Let $P$ (e.g., $P = 2$) students represent a
packet. The sender node \texttt{A} ``transmits'' the packet by sending
off the students at a rate of $R$ students per second (e.g., $R = 1$
student per second). A student then walks along the communication link
of length $l$~meters (e.g., $l = 10$~meters), which can be marked on
the ground of the PE field at a speed (pace) of $s$ (e.g., roughly $s
= 1$~meter per second). When the first student arrives at the first
intermediate node, the student has to wait for the remaining $P - 1$
students to form the complete ``packet'' at the first intermediate
node.

\begin{figure}[t!]  \centering
\includegraphics[width=3.5in]{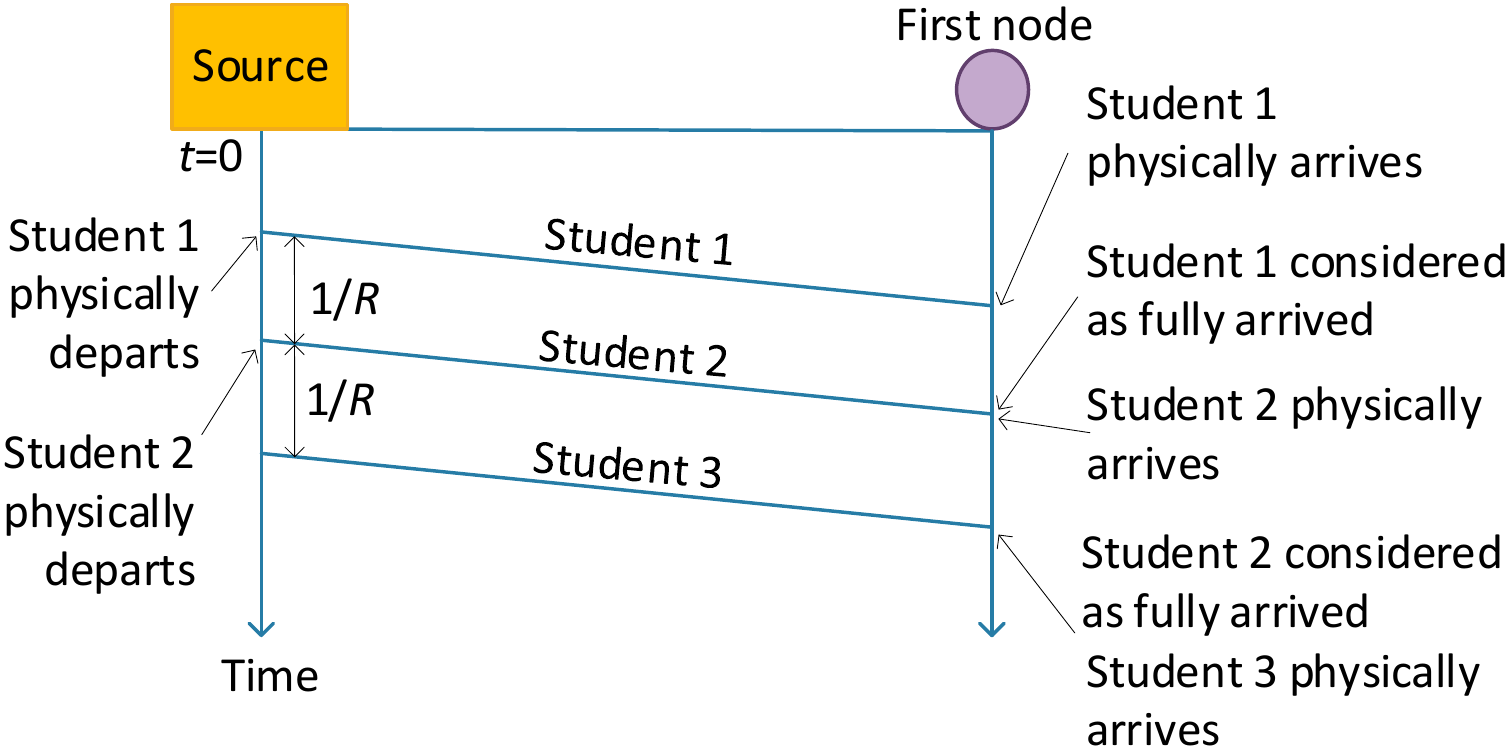}
\caption{Illustration of students moving from a source to a first
network node to represent the transmission of individual bits of
information of a given traffic flow. With transmission rate
$R$~bit/second, there is a student ``transmitted'' (dispatched) from
the source every $1/R$~seconds. In a real network, the transmission
of one bit takes $1/R$~seconds, i.e., the transmission of one bit is
a ``process'' that takes a duration of $1/R$~seconds. Thus, even
though Student $1$ physically departs at time zero, the transmission
of the one bit that Student 1 represents is completed only at time
$1/R$ (when the transmission of Student $2$ (bit $2$) commences);
accordingly, Student $1$ is considered as fully arrived at the first
node when Student $2$ physically arrives to the first node.}
\label{student_tra:fig}
\end{figure}
More specifically, with a student
transmission rate of $R$ students per second, it takes $1 \mbox{
student} / R \mbox{ students/s} = 1/R \mbox{ s}$ to ``transmit'' one
student. That is, the transmission of the first student is from a
mathematical perspective only complete when the second student
physically departs at time $1/R \mbox{ s}$, even though the first
student physically departed at time 0, i.e., at the start of the
transmission process.  Accordingly, the first student physically
arrives at the first intermediate node at time $l/s$; however, from a
mathematical perspective, the reception is only complete after $1/R +
l/s$.  Essentially, the basic rule for the modeling of the bit
transmission with student movements is that a student is considered to
have fully arrived exactly $1/R$~s (i.e., the time spacing between two
successive student transmissions) after the physical arrival at the
next node.  Practically, a student can consider herself/himself as
fully arrived when the next student physically arrives to the node, as
illustrated in Fig.~\ref{student_tra:fig}. Alternatively, a student
can consider herself/himself as fully arrived after waiting for the
time span $1/R$, e.g., $1$ second for $R = 1$~bit/s, after physically
arriving to a node.

When all $P$ bits forming a packet have completely arrived, i.e., when
the next bit, i.e., bit number $P+1$ arrives, the first
intermediate node sends the $P$ students (the packet) off at the
transmission rate $R$ and the students propagate to the second
intermediate node. This process repeats until the complete ``packet'',
i.e., all $P$ students, arrive to the destination node \texttt{B}.
Again, the $P$th student is considered as fully arrived exactly
$1/R$~s after the $P$th student physically arrives.

As a simplification, if this distinction between physical arrival and
``being considered as fully arrived'' is too complicated for the
students or would cause undue confusion, then this distinction can be
neglected. This distinction becomes mathematically negligible when the
packet size $P$ becomes large, since for large packet size $P$, there
is only a small difference between considering physical arrival,
resulting in a delay of $(P-1)/R$~s for transmitting the packet to the
next node, and considering the ``full arrival'', resulting in a delay of
$P/R$~s.

This distinction between physical arrival of a student and the student
being considered as fully arrived is due to the physical differences
between sending a discrete entity (e.g, a student) and sending one bit
of information. A bit transmission requires a prescribed signal (e.g.,
electrical signal or optical signal) to be sent into the communication
link (e.g., wire of optical fiber) over a time period of $1/R$~s,
i.e., the time duration that it takes to transmit one bit of
information.

The teacher or outreach assistants should measure the time from the
time instant when the first student departs the sender node \texttt{A}
to the time instant of the arrival of the $P$th student to the
destination node \texttt{B}.

\begin{figure}[t!]  \centering
\includegraphics[width=3.5in]{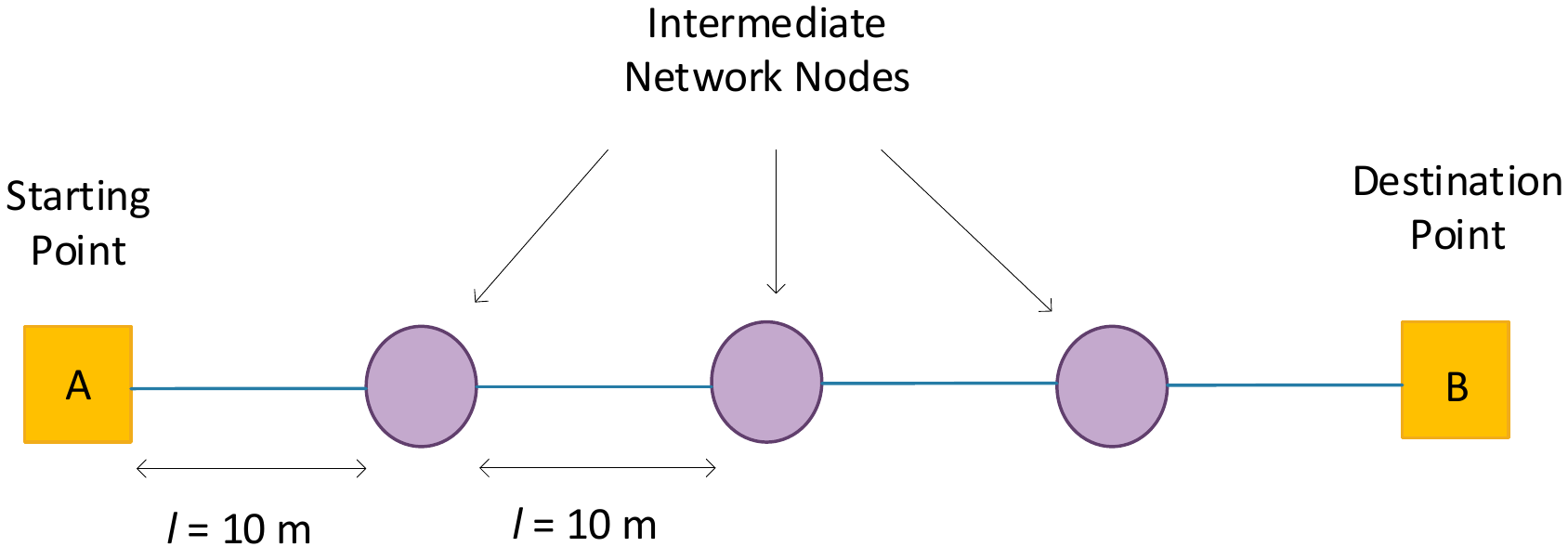}
\caption{Example network with a source (starting point), three
intermediate network nodes, and a destination point
(destination). Each of the four communication links spans a physical
distance of $l = 10$ meters.}
\label{netw:fig}
\end{figure}
Based on the described modeling of the store-and-forward packet
switching the students can be tasked with a variety of races where
student teams race against each other to deliver a message of $M$ bits
from \texttt{A} to \texttt{B} while following the rules of
store-and-forward packet switching.  The teacher or outreach
coordinators need to set up two separate ``networks''. Each network
consists of a starting point \texttt{A}, several intermediate network
nodes, say $N = 3$ intermediate network nodes, a destination node
\texttt{B}, see Fig~\ref{netw:fig}.
There is a distance of say $l = 10$~meters between two
nodes, i.e., a complete network spans $(N +1) l = 40$~meters.
Let each link operate with a transmission bitrate of $R = 1$~bit/s
and propagation speed $s = 1$~m/s.

\paragraph{Message vs. Packet Switching Race}
For a typical class of say $24$ students, a message-vs-packet switching
race can be conducted as follows: Divide the class into two groups of
$12$ students each. Each group of $12$ students represents a message of $M
= 12$~bits, e.g., a text message. Then assign one group to message
switching where the entire message of $M = 12$~bits is treated as one
``packet''. That is, all $M$ bits (students) need to be assembled at
each intermediate node before the message can be forwarded to the next
node. Assign the other group to packet switching with a small packet
size, e.g., $P = 3$~bits (students). For packet switching, only the
$P$ bits need to be assembled at an intermediate node before the
packet can be forwarded to the next node. Ask the students about their
expectation as to which group will finish faster, or
whether both groups will take the same time?

\begin{figure*}[t!]  \centering
\includegraphics[width=5in]{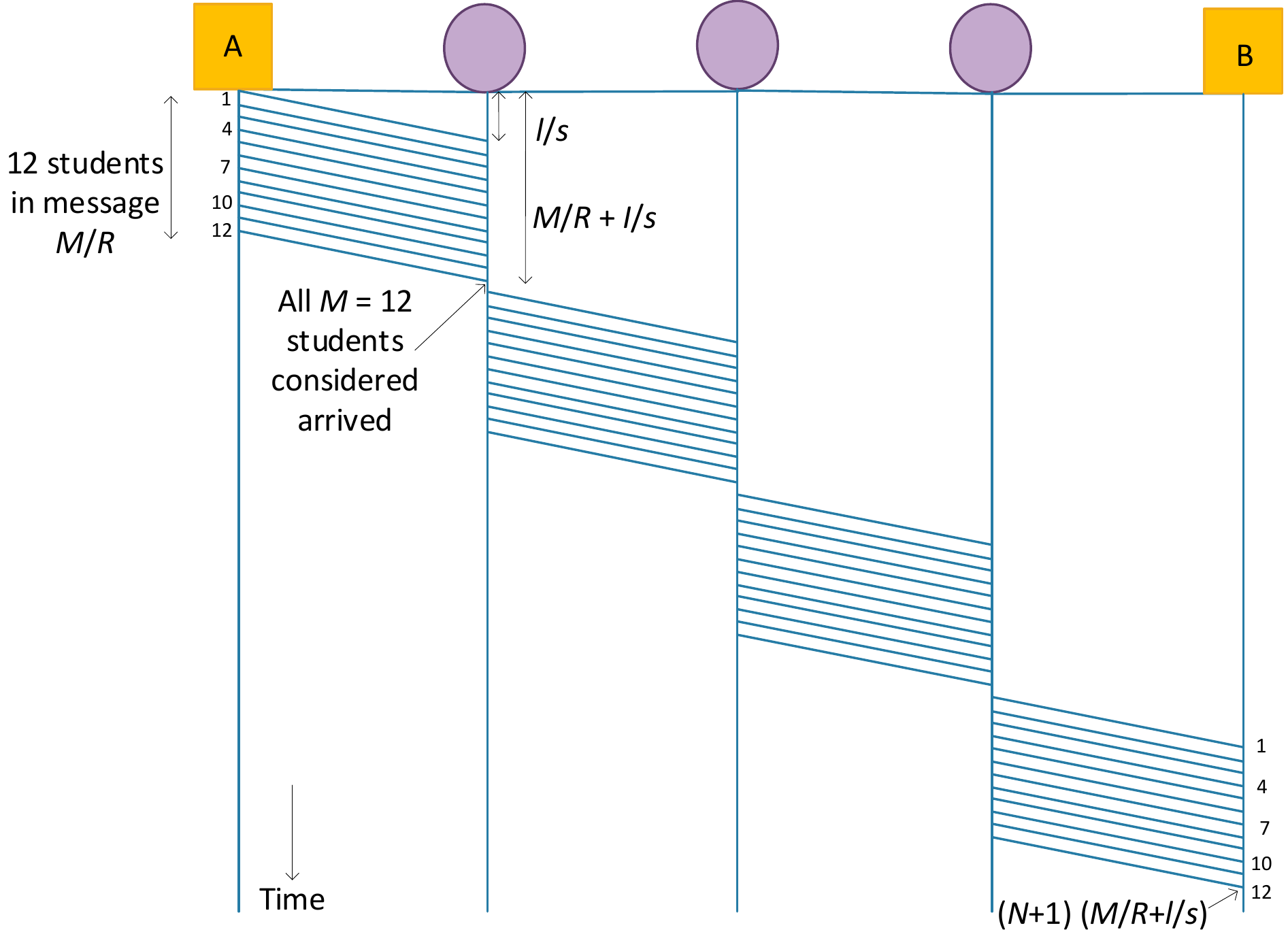}
\caption{Illustration of timing dynamics of transmission of a message
consisting of $M = 12$~bits (students) with message switching from
source~A to destination~B via three intermediate network nodes. Each
intermediate network node needs to receive the entire message (i.e.,
all $M$~bit) before the message can be processed and forwarded to
the next network node.}
\label{mess:fig}
\end{figure*}
Let the students start at the same time to ``transfer'' the message
from \texttt{A} to \texttt{B}.  Observe how the group with message
switching transfers bits only over one link at a time, see
Fig~\ref{mess:fig}. In contrast, once the transmission process has
gotten underway, the group with packet switching transfers bits over
multiple links, i.e., exploits a so-called pipelining effect.
Accordingly, the message switching group will take much longer to
transfer the message from \texttt{A} to \texttt{B}.  In particular,
the message switching group will incur a total delay of
\begin{eqnarray}
  (N+1) \left[ \frac{M}{R} + \frac{l}{s} \right],
\end{eqnarray}
which gives for the
considered example
\begin{eqnarray}
  (3+1) \left[ \frac{12 \mbox{ bit}}{1 \mbox{ bit/s} }
    + \frac{10 \mbox{ m}}{ 1 \mbox{ m/s} } \right ] = 88 \mbox{ seconds}.
\end{eqnarray}
When neglecting the distinction between physical arrival of a student
and the student being considered as fully arrived,
then the total delay for the simplified version of
considering physical arrivals is
\begin{eqnarray}
  (N+1) \left[ \frac{M-1}{R} + \frac{l}{s} \right], i.e.,
\end{eqnarray}
\begin{eqnarray}
  (3+1) \left[ \frac{12 - 1 \mbox{ bit}}{1 \mbox{ bit/s} }
    + \frac{10 \mbox{ m}}{ 1 \mbox{ m/s} } \right ] = 84 \mbox{ seconds}.
\end{eqnarray}

\begin{figure*}[t!]  \centering
\includegraphics[width=6in]{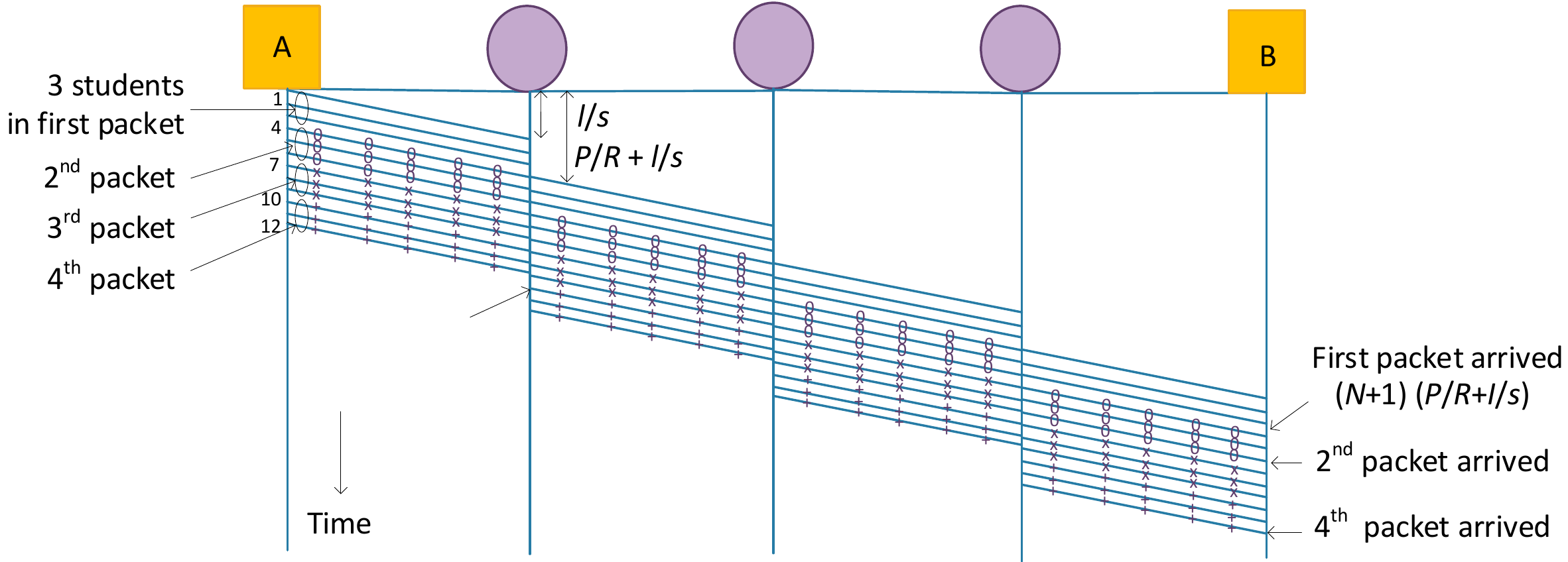}
\caption{Illustration of timing dynamics of transmitting a message
consisting of $M = 12$~bit that is partitioned into four packets,
each containing $P = 3$~bit. An intermediate node needs to receive
all $P$ bit of a packet before the packet can be processed and
forwarded.}
\label{pkt:fig}
\end{figure*}
In contrast, for the packet switching group, the first packet
arrives after
\begin{eqnarray}
  (N+1) \left[ \frac{P}{R} + \frac{l}{s} \right],
\end{eqnarray}
which gives for the considered example
\begin{eqnarray}
  (3+1) \left[ \frac{3 \mbox{ bit}}{1 \mbox{ bit/s} }
    + \frac{10 \mbox{ m}}{ 1 \mbox{ m/s} } \right ] = 52 \mbox{ seconds}.
\end{eqnarray}
Then, every $P/R = 3 \mbox{ bit} / 1 \mbox{ bit/s} = 3$~seconds,
another packet arrives to \texttt{B}, see Fig.~\ref{pkt:fig}.
Since there are three more packets needed to make the
message complete at \texttt{B}, the total
delay for the packet switching group is $52 + 3 \times 3 \mbox{ s} =
61$~seconds, i.e., $17$ seconds before the message switching group
finishes.
Or, when considering only the physical arrival of students,
\begin{eqnarray}
  (N+1) \left[ \frac{P-1}{R} + \frac{l}{s} \right],
\end{eqnarray}
i.e.,
\begin{eqnarray}
  (3+1) \left[ \frac{3 - 1 \mbox{ bit}}{1 \mbox{ bit/s} }
    + \frac{10 \mbox{ m}}{ 1 \mbox{ m/s} } \right ] = 48 \mbox{ seconds}.
\end{eqnarray}

The next question for the students is then how the delay can be
minimized? For our setting, which ignores the packet header size and
operates with a granularity of individual students, i.e., bits, the
minimum delay is attained for a packet size of $P = 1$:
\begin{eqnarray}
  (3+1) \left[ \frac{1 \mbox{ bit}}{1 \mbox{ bit/s} }
    + \frac{10 \mbox{ m}}{ 1 \mbox{ m/s} } \right ] = 44 \mbox{ seconds}
\end{eqnarray}
to get the first packet to \texttt{B}. Then eleven more packet
(students), arriving at a rate of $R = 1$~student per second into
\texttt{B}, add another $11$~seconds, for a total delay of $44 + 11 =
54$~seconds.

As an alternative, consider a faster-paced message switching versus
packet switching race with a propagation speed $s = 2$~m/s.  The
message switching group will incur a total delay of
\begin{eqnarray}
  (3+1) \left[ \frac{12 \mbox{ bit}}{1 \mbox{ bit/s} }
    + \frac{10 \mbox{ m}}{ 2 \mbox{ m/s} } \right ] = 68 \mbox{ seconds}.
\end{eqnarray}
For the packet switching group, the first packet
arrives after
\begin{eqnarray}
  (3+1) \left[ \frac{3 \mbox{ bit}}{1 \mbox{ bit/s} }
    + \frac{10 \mbox{ m}}{ 2 \mbox{ m/s} } \right ] = 32 \mbox{ seconds},
\end{eqnarray} and
every $P/R = 3 \mbox{ bit} / 1 \mbox{ bit/s} = 3$~seconds the next
packet arrives, for a total delay of $32 = 3 \times 3 \mbox{ s} =
41$~seconds, i.e., $27$ seconds before the message switching group
finishes.

\subsubsection{HTML Web Page Download}
\paragraph{Background}
The download of a web page consisting of a hypertext markup language
(HTML) base page with embedded (linked) objects using the hypertext
transfer protocol (HTTP) is the basis for web browsing.  The HTTP
protocol is an application layer protocol that operates over the
reliable transmission control protocol (TCP) transport layer
protocol. Accordingly, a TCP connection must be established between a
client and the web server before the web page can be requested from
the web server. Commonly, the HTTP request message is piggybacked on
the third part of the three-way TCP handshake~\cite{for2012com,kur2017com}.
In order to evaluate the web
page download delay, let $RTT$ denote the round-trip delay from the
client to the web server and back to the client in seconds. The $RTT$
typically represents the delay for sending a small packet and thus
characterizes the propagation delays, queuing delays, and processing
delays on the path from the client to the server and back to the
client. Let $b$ denote the size of the HTML base page in bit and $o$
denote the size of an embedded object in bit. Suppose that there is
one link between client and server with transmission bit rate
$R$~bit/s (in both directions).

\begin{figure}[t!]  \centering
\includegraphics[width=3.5in]{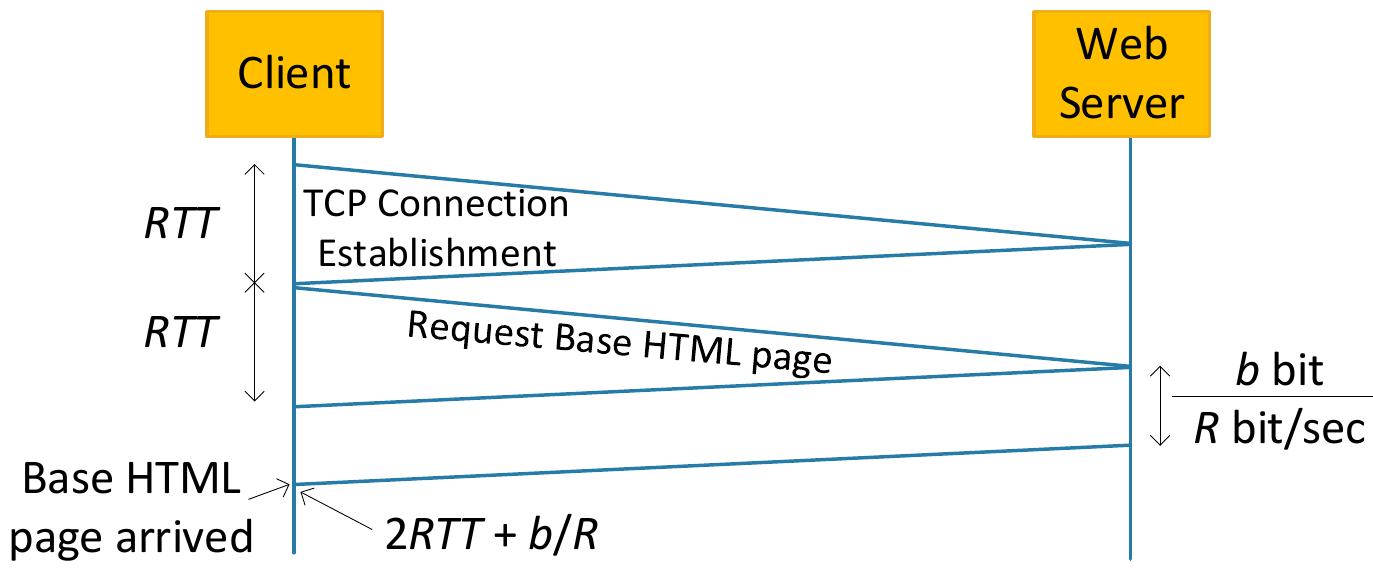}
\caption{Illustration of timing dynamics of downloading an HTML base
  web page from a web server. The establishment of the TCP connection
  incurs one round trip time $RTT$ delay, followed by one round trip
  time $RTT$ for the actual base page request and propagation to the
  client; in addition, the transmission of the $b$ bit of the base
  page incurs a transmission delay of $b/R$.}
	\label{htmlba:fig}
\end{figure}
The download of the HTML base page requires $2 RTT +
b/R$, see Fig~\ref{htmlba:fig}.  With the default non-persistent
HTTP connections, the web server closes down the TCP connection after
sending the base page to the client. Upon receipt of the base page,
the web browser at the client parses the base page and discovers the
links to the embedded objects. The browser then needs to request each
of the embedded objects from the web server.  A basic approach is to
use non-persistent connections and to request one embedded object at a
time. A new TCP connection needs to be established via the $3$-way TCP
handshake for each embedded object, leading to a delay of $2 RTT +
o/R$ for each embedded object. This sequential requesting of one
embedded object at a time results for $E,\ E \geq 1$, embedded objects
in a delay of $E [2 RTT + o/R]$.  The total web page down load delay is
thus
\begin{eqnarray}  \label{wpdnop:eqn}
  2 RTT + \frac{b}{R} + E \left[ 2 RTT + \frac{o}{R} \right].
\end{eqnarray}

Alternatively, the browser can use up to $C$ parallel connections, each
with the same round trip time $RTT$ and bitrate $R$.
With $C$ parallel connections, the $E$ embedded objects can be retrieved
with $E/C$ ``rounds'' of establishing $C$ parallel TCP connections and
transmitting $C$ embedded objects in parallel (simultaneously) from
the web server to the client. Assuming that $E/C$ is an integer,
the total download delay is
\begin{eqnarray}
  2 RTT + \frac{b}{R} + \frac{E}{C} \left[ 2 RTT + \frac{o}{R} \right].
\end{eqnarray}

\paragraph{Student Modeling}
The HTTP download interactions between client and web server can be
modeled through students running back and forth between a client
location (position) on a field and a server location that is some
distance, i.e., $l$ meters away. If students propagate (walk or run)
on the link between client and web server at a speed of $s$~m/s, then
the round trip time is $RTT = 2 l/s$, e.g., for a distance of $l =
15$~m and running speed $s = 3$~m/s, the round trip time is $RTT =
10$~s.

The client node and the web server can be represented by two students
that are stationary (providing an opportunity for the involvement of
mobility-limited students). The packets required for TCP connection
establishment as well as the base object and the embedded objects can
be represented through students. More specifically, one ``TCP
connection establishment'' student runs from the client to the server
and back to the client to ``establish'' the TCP connection. If
desired, the established connection can be indicated through a rope
that the student drags back from the server to the client. The rope on
the ground represents the established TCP connection.  Then, the same
student, or a different ``HTTP request'' student runs along the rope
from the client to the server to carry the request for the HTML base
page to the server. The base page can be represented through $b$
``base page'' students.  More specifically, the HTML base page can be
represented by a sheet of paper that lists the embedded objects, e.g.,
Image~$1$ of size $o_1$~bit, Image~$2$ ($o_2$ bit), Video~$1$ ($o_3$
bit). The base page sheet can be cut up into $b$ (vertical) slices and
each base page student carries one slice.  The base page students are
dispatched at the transmission rate of $R$ bits (students) per
seconds.  Once all $b$ students have arrived at the client, they give
their slices of the base page sheet to the client, who assembles the
complete sheet and begins to request the embedded objects. While the
client assembles the sheet, the server pulls in the rope to signify
the closing down of the TCP connection, following the standard
non-persistent HTTP protocol.

The client then sends a TCP connection establishment student to the
server to establish a new TCP connection, i.e., to pull the rope from
the server to the client.  Then, an ``HTTP request'' students is sent
to the server to request the first embedded object.  The $o_1$
``embedded object~$1$'' students are then dispatched by the server along
the rope to the client at a rate of $R$ bits (students) per second.
Once all $o_1$ students have arrived at the client, the server pulls
in the rope and the client starts the process of requesting embedded
object $2$ by sending a ``TCP connection establishment'' student to the
server. This process of establishing a TCP connection, requesting an
embedded object, and closing down the TCP connection is repeated for
each embedded object.

\paragraph{Specific Basic Example Without Parallel Connections}
Consider a class with $24$ students. Assign one student to be the client
and one to be the server. The client needs one TCP connection
establishment student and one HTTP request student (that runs back
with the objects from the server to the client).  Let $b = 3$ students
represent the HTML base page and $o = 6$ students represent one
embedded object, whereby there are $E = 3$ embedded objects in the web
page.  The goal of the class is to get the base page and the three
embedded objects from the server to the client so that the client can
display the full web page.  In this example, with a round trip time
$RTT = 10$~s (distance $l = 15$~m, running speed 3~m/s) and bitrate $R
= 1$~bit/s, the total time for downloading the web page should be as
per Eqn.~(\ref{wpdnop:eqn}) equal to
\begin{eqnarray}
  2 \times 10 \mbox{ s} + \frac{3 \mbox{ bit}}{1 \mbox{ bit/s}}
  + 3 \left[ 2  \times 10 \mbox{ s}
    + \frac{6 \mbox{ bit} }{ 1 \mbox{ bit/s}} \right]  = 101 \mbox{ s}.
\end{eqnarray}

\paragraph{Web Page Download Race with Parallel Connections}
Once a class is proficient in performing the ``web page download''
without parallel connections, it is ready for racing other classes or
groups from one class can race against each other.  To make the races
interesting, it is recommended to operate with different numbers $C$
of parallel connections (or race a group with, say $C = 2$ parallel
connections versus a group without parallel connections).  Parallel
connections allow the client to open multiple ($C$) TCP connections in
parallel to the server to download $C$ embedded objects in parallel.

Consider a modified version of the specific example with $E = 6$
embedded object, each of size $o = 3$ bit and $C = 2$ parallel
connections.  After the client obtains the HTML base page, the client
sends two TCP connection establishment students simultaneously
(side-by-side) to the server, returning with two parallel rope lines
to the client.  The client can then simultaneously send two HTTP
request students along the two rope lines to the server.  The server
then sends two embedded objects, i.e., two groups of $o = 3$ students
each, in parallel onto the two rope lines.  The transmission rate is
$R = 1$ student per second on each rope line (TCP connection).  Once,
the $o = 3$ students arrive in parallel to the client, the two rope
lines are pulled back by the server and the client sends out the next
two TCP connection requests in parallel. With this parallel download
of the embedded objects, all $E = 6$ embedded objects can be
downloaded in $E/C = 6/2 = 3$ rounds.  The total delay is thus:
\begin{eqnarray}
  2 \times 10 \mbox{ s}  + \frac{3 \mbox{ bit} }{ 1 \mbox{ bit/s}}
  + \frac{6}{2} \left[ 2 \times 10 \mbox{ s}  + \frac{3 \mbox{ bit}}{1 \mbox{ bit/s}} \right] = 92 \mbox{ s}.
\end{eqnarray}
The students can be challenged to figure out the protocol parameters
that minimize the download delay. A higher number $C$ of parallel
connections will reduce the download delay. In the ``browser wars'',
competing web browsers pushed the number $C$ of parallel connections
higher and higher to achieve shorter delays than their
competitors. Higher numbers $C$ of parallel connections increase the
complexity of the web browser and the management overhead on the
operating system. Eventually, most major browsers agreed to a
``truce'', setting the number $C$ of parallel connections in the range
of seven to ten.

\paragraph{Extension to Caching}
The students can be further challenged to devise ways to reduce the
web page download delay. As the students will notice in the races
outlined above, a substantial amount of time is spent running back and
forth between client and server. Can this ``running time'', i.e., the
round trip time $RTT$ be reduced?

The answer is ``Yes'', and the solution approach lies in the typical
request pattern of a group of students, e.g., at a school. Likely, all
the students at the school regularly access the web page of the
learning management system employed at the school, or specific web
pages that are recommended by teachers. Accordingly, there are a few
popular web pages that are requested frequently. These popular web
pages or at least the numerous embedded objects in these popular web
pages can be cached (i.e., stored on a computer with a large memory)
near the school. This local caching will reduce the $RTT$ for these
embedded objects. The base HTML page can still be obtained from a
distant server to get the latest updates to the web page, while the
embedded objects change typically less frequently and can be cached.

The student model of the web page download can be modified as follows
for caching. A local cache node is placed in the vicinity of the
client, e.g., at a distance $l_{\rm cache} = 3$~m from the
client. After the client has obtained the base HTML page from the
still distant (e.g., $l = 10$~m) server, the embedded objects are
obtained from the nearby ($l_{\rm cache} = 3$~m) cache.  The short
round trip time $RTT = 2 \times l_{\rm cache} / s = 2 \times 3 \mbox{
  m} / 3 \mbox{ m/s} = 2$~s to the local cache substantially reduces
the download delay.  Content distribution companies, such as Akamai,
as well as large web content providers, e.g., Amazon and Netflix, have
installed caches close to our local neighborhoods and are thus able to
provide us with low-latency Internet services and content.

This caching can be taken a step further in wireless networks with
several smart phones that are near each other, e.g., when a class or a
group of friends meet, e.g., during school recess or passing
period. Content can be cached on the smart phones and thus, we carry
the content essentially around with us. When a nearby friend wants to
watch a video clip that you have just watched, then the friend's smart
phone can request the video clip from the cache inside your
phone. This way, friends---or rather, the smart phones of
friends---can help each other to provide low-delay streaming services.

\subsubsection{Medium Access Control: LTE Connection Establishment}
\paragraph{Background}
Smart phones need to connect with a cellular wireless network or a
WiFi network to provide Internet services. The connection to a
wireless cellular network is needed when there is no local WiFi
network to connect to.  The connection establishment to a cellular
network follows a medium access control (MAC) protocol, e.g., the Long
Term Evolution (LTE) MAC protocol in $4$G wireless
networks~\cite{ami20143gpp,de2017random,lay2013ran,tya2013imp,tya2015con,vil2017lat}.
In this MAC protocol, a smart phone sends out a so-called preamble,
which is essentially a connection establishment request. The
connection establishment system is set up with a limited number of
available slots into which smart phones can sent their requests
(preambles). Typically, there are many smart phones within the reach
of a particular cellular network (cell tower). These smart phones (and
their users) are all geographically distributed and they do not
coordinate among each other before starting a connection
request. Therefore, the requests arrive into the slots of cellular
system as a random process.  Also, there can be many (possibly
thousands or more) smart phones in the reach of a cell tower, but
there are only few slots, say $S = 8$ slots, so it is not possible to
statically assign a slot to each smart phone that could possibly be
near a cell tower. Rather, the LTE MAC protocol requires that each
smart phone sends its request into a randomly chosen slot. The
transmission of the requests proceeds in rounds. If a slot contains
only one request in a round, then this request is successful. If a
slot contains two or more requests, then these requests collided and
are not successful; these unsuccessful requests need can be
re-transmitted in a future round.

\paragraph{Student Modeling}
We can model the LTE connection establishment with a variation of the
musical chairs game. Let there be $S$ chairs that represent the slots
for the LTE connection requests (preambles).  Suppose there are $P,\ P
> S$, smart phones that need to get connected, e.g., $P$ could be
equal to the number of students in the class.  In the first round, all
$P$ students participate in the musical chairs game with $S,\ S < P$,
chairs. When the music stops, the chairs with one student represent
successful slots with one LTE connection request. These students who
were alone sitting on a chair can go to a ``connected'' section of the
room. The remaining students participate in the next round of the
game.  The goal is to get all students ``connected'', i.e., each
student continues participating in the successive rounds until s/he
sits alone on a chair and can then go to the ``connected'' status.
For good phone service, we expect that the connection establishment
is completed quickly. The number of rounds until all students are
``connected:'' should be as low as possible so that all phones are
connected quickly.

A class of, say $24$ students can be split into two groups of $12$
students each that compete against each other to connect every
student in the group in the fewest number of rounds.  How can the
number of required rounds be reduced? In a real wireless system, the
smart phones usually cannot coordinate among each other, after all,
they first need to be connected before they can communicate with the
outside world, and exactly this process of getting connected would be
helped with some coordination. In our classroom modeling of the MAC
protocol, we can experiment with coordination among the phones
(students) in a group.  How should the students in a group play the
game to reduce the number of required rounds? Suppose that there are
$P = 12$ students and $S = 4$ chairs (slots). What is the best
coordination strategy?

The students can coordinate such that three students sit individually
on three chairs and all remaining nine students go to the fourth
chair. Thus, $S - 1 = 3$ phones (students) are successful in the first
round.  Following this strategy, three more students can be successful
in each successive round, requiring a total of four rounds.

An alternative strategy that is used in real wireless networks and
works without coordination is to bar some students from participating
in the rounds. If there are initially $12$ students, we could initially
bar six students from participating and let only the remaining six
students go into the first round of the musical chairs game (and these
initial six play without coordination). After all the initial six
students have been connected, we let the other six into the game. The
students can experiment with initially barring different numbers of
students and with different strategies for adding initially barred
students into the game. These experiments can be done by two groups in
parallel as a race that is won by the group that gets all students
connected in the fewest number of rounds.

\subsubsection{SDN Control Plane and Data Plane Interactions:
Data Packet Routing} This section outlines the development of a PE
race activity, the SDN Networking Race, that brings the SDN-based
network operation to life in a team relay race. The SDN Networking
Race emulates the operation of an SDN network by having student
runners embody the data and control packets.  We outline race
scenarios that require control packet runners to reach a prescribed
``central'' controller location and retrieve an instruction sheet to
represent the controller flow action. The control packet runner needs
to return with the flow action sheet to the originating switch
location before data packet runners commence their run through a
prescribed route of distributed switch locations. This SDN mode of
operation with a centralized control is pitted against classical
destination-based IP routing to illustrate the tradeoffs in
IP vs. SDN routing.

\paragraph{Background}
Classical Internet Protocol (IP) packet routing is only based on the
destination address of a packet. That is, an IP router only considers
the IP destination address in the header of an IP packet to decide how
to route (forward) the packet. The IP packet routers are
pre-configured with the destination address based routes, i.e., a
router can immediately decide how to forward an incoming IP packet.

In contrast, the routing functionality of SDN packet switches is
controlled by a central controller~\cite{aky2014roa,guc2016fun,kot2012out}.
In an abstract sense, the SDN switches that handle the actual
data packet routing form the ``data plane''. On the other hand,
the controller that controls the SDN switches forms the ``control plane''.
The interactions between the SDN control plane and data plane enable a
wide range of optimization mechanisms that enhance the provisioning of
services to users~\cite{kar2020mul,wan2019mul}.

When a data packet of a new data flow arrives to an SDN switch, i.e.,
the SDN switch receives a packet for which there are currently no
action rules configured on the switch, then the switch asks the
controller what to do with this packet. The central controller has the
overview over the entire network of switches and can devise some
optimized routing action for the new data flow.  The controller sends
the routing action rules for this new data flow to all involved
switches, i.e., configures the switches along the routing path of the
data flow. This process of the switch asking the controller before
actually forwarding the packet costs some time, i.e., the round trip
time from the switch to the controller and back to the switch.
However, with the optimized flow routing that the controller devised,
the data flow can hopefully be forwarded more efficiently and make up
the initial delay for asking the controller. Thus, there is a chance
that the SDN mode of operation is faster in the end.  We outline next
an SDN Networking Race to see what is faster, IP routing or SDN
routing.

\paragraph{Network Layout and Traffic Flows}
\begin{figure}[t!]  \centering
\includegraphics[width=3in]{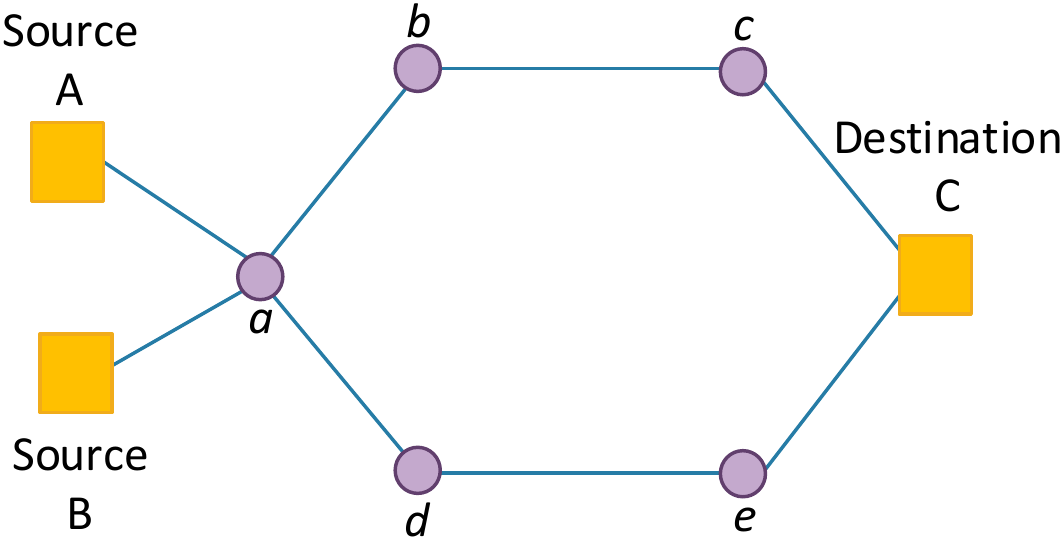}
\caption{Network topology for SDN Networking Race: Two source nodes
  \texttt{A} and \texttt{B} traffic flows to a destination node
  \texttt{C} via a network with intermediate network nodes $a$, $b$,
  $c$, $d$, and $e$.}
	\label{routenet:fig}
\end{figure}
Consider the network with two sources (\texttt{A} and \texttt{B}), one
destination (\texttt{C}), and a network of intermediate network nodes
($a$, $b$, $c$, $d$, and $e$) as illustrated in Fig.~\ref{routenet:fig}.
This network topology should be marked twice
on the PE field, e.g., with chalk. That is, there should be an IP
version of the network in Fig.~\ref{routenet:fig} and an SDN version
of the network in Fig.~\ref{routenet:fig}. The networks should be
sufficiently large and for fairness need to be of exactly the same
dimensions, so that the students running from the sources via the
intermediate network nodes have some distance to cover and each of the
intermediate network nodes has enough space on the ground to draw out
the queue, see Fig.~\ref{node:fig}.

For classical IP routing, suppose that the network is configured such
that intermediate node $a$ routes all packets with destination address
\texttt{C} to the intermediate node $d$, while $d$ routes to $e$, and $e$
routes to \texttt{C}. Notice that with classical IP routing all
traffic towards \texttt{C} is routed along the ``bottom'' route in
Fig.~\ref{routenet:fig}, i.e., $d$ and $e$, while the ``top'' route, i.e.,
$b$ and $c$ are idle.

For SDN routing, the controller can look at the entire network and
notice that the traffic flows arriving into $a$ come in from two
different sources, namely \texttt{A} and \texttt{B}. The SDN
controller can thus optimize the routing, e.g., by routing the traffic
flow from \texttt{A} to \texttt{C} over the top route, i.e., $b$ and $c$,
while the traffic flow from \texttt{B} to \texttt{C} is routed over
the bottom route, i.e., $d$ and $e$.

Let all communication links in the network have the same transmission
bitrate $R$, e.g., $R = 1$~packet/s (for some arbitrary packet size)
and the same propagation speed $s$.

\paragraph{Student Modeling}
In order to enforce the proper ``network operation'', it is
recommended that at least the sources \texttt{A} and \texttt{B} as
well as the first intermediate node $a$ are ``staffed'' with teachers or
outreach coordinators to enforce the transmission bitrate $R$ and to
remind students to walk or run with the same pace $s$ from one node to
the next. If there is enough personnel, then the other nodes $b$,
$c$, $d$, and $e$ should also be staffed; alternatively, these nodes
could be monitored by the staff at the other nodes.

To set up the SDN Networking Race, split a class in half, e.g., a
class of $24$ students into two groups of $12$ students each, i.e., an IP
group of $12$ students, and an SDN group of $12$ students.  Then split
each group in half and position six students each at source \texttt{A}
and six students at source \texttt{B}. The goal is to get all 12
students in a group to the destination \texttt{C} as quickly as
possible. Each student should be given a sheet of paper with the
source (\texttt{A} or \texttt{B}) and the destination (\texttt{C}),
this sheet models the packet header.

Throughout, this SDN Networking Race should preferably be conducted
with the distinction that a student is considered only fully arrived
at a node when the next student arrives to the node, see
Section~\ref{stumodps:sec}. This is because we let each student
represent one data packet in this race and we need each student to
model a finite transmission delay of packet size divided by
transmission bitrate in order to model store-and-forward packet
switching of whole packets. Practically, this can be explained to the
students that each student needs to wait for $1/R$, e.g., one second
for a transmission rate of $R = 1$~packets/s, after arriving at a node
before becoming eligible for onward transmission.  Note that in this
race two packet traffic flows mix at node $a$; whereas, in the
store-and-forward activity in Section~\ref{stoforps:sec} only a single
traffic flow traversed the network. In Section~\ref{stoforps:sec},
specifically, in Fig.~\ref{student_tra:fig}, a student could be
considered as fully arrived at a node, when the next student arrived
at the node. This ``looking out for the next student'' to determine
the waiting time at a node worked in Fig.~\ref{student_tra:fig}
because all students belonged to the same traffic flow. However, in
the SDN Networking Race, the students arriving to an intermediate
network node may belong to different traffic flows and packets from
the different traffic flows may get interspersed with each
other. Therefore, this looking out for the next student no longer
works in the SDN Networking Race; rather students need to actually
measure time to determine the $1/R$ waiting time at a node until they
can be considered as fully arrived.

Alternatively, if this waiting for $1/R$ is too complex for the
students, then the SDN Networking Race can be conducted with each
student immediately passing through a switching node and not waiting.
However, each intermediate node can still only transmit at rate
$R$~students per second. Thus, if an intermediate node has not
transmitted a packet in some time (in over $1/R$ specifically), then
an arriving student can immediately pass through and proceed to the
next node along the prescribed route.  On the other hand, if a node
has just transmitted a student and a new student arrives at this same
time instant, then the newly arrived student has to wait for $1/R$
before it can be transmitted. This behavior models the so-called
``cut-through'' switching functionality of some switches.

For both the IP network and the SDN network, the sources \texttt{A}
and \texttt{B} send students at the transmission bitrate $R$, e.g., $R
= 1$~student per second, into the network.  In particular, source
\texttt{A} sends one student per second to intermediate node $a$ and
source \texttt{B} also sends one student per second to intermediate
node $a$. For IP routing, the intermediate node $a$ is an IP router
and only looks at the destination address, which is \texttt{C} for all
packets (students). According to the preconfigured routing policy,
the router sends all students onwards to $d$ at the transmission rate
of $R = 1$~student per second.  The students in the queue at node $a$
are served in a first-come-first-served manner, i.e., newly arriving
students (from \texttt{A} or \texttt{B}) join the back of the queue.
The intermediate node $d$ forwards the students to $e$ at $R =
1$~student per second and intermediate node $e$ forwards to \texttt{C}
at $R = 1$~student per second.

With SDN, the first packet of the \texttt{A} to \texttt{C} traffic
flow triggers a query to the controller. This query can be modelled by
letting the staff member at $a$ run to a controller location that is
some distance away to retrieve the optimized routing policy (flow
routing action) that the \texttt{A} to \texttt{C} traffic flow should
be routed via $a$, $b$, and $c$.  The intermediate node $a$ can only
start forwarding the first packet to intermediate node $b$ when the
routing policy (represented by a sheet of paper with the routing
policy written on it) has arrived back at $a$ via the running staff
member.  Similarly, the arrival of the first packet (student) of the
\texttt{B} to \texttt{C} traffic flow triggers another query to the
controller, i.e., the staff member at $a$ has to run to the controller
once more to retrieve the routing policy via $a$, $d$, and $e$ for the
\texttt{B} to \texttt{C} traffic flow. Once the staff member of
intermediate node $a$ is back at node $a$, both traffic flows can be
forwarded.  Importantly, node $a$ has two outgoing links, one link to
$b$ and one link to $d$, and each link operates with transmission rate
$R$. Thus, node $a$ can send the traffic flow \texttt{A} to \texttt{C}
at a rate of $R = 1$~student per second to node $b$, and in parallel
send the traffic flow \texttt{B} to \texttt{C} at a rate of $R =
1$~student per second to node $d$.

The students will observe that the longer the ``transmission'' of the
packets goes on, the more and more the SDN network will make up the
initial delay for obtaining the routing policy from the controller and
can ultimately get all packets faster to the destination \texttt{C}.
The students can be tasked with figuring out variations of the SDN
Networking Race, e.g., how can the SDN network be made to work faster?
Clearly, if the controller is close to the first node $a$, or a very
fast runner is employed to retrieve the routing policies from the
controller, then the initial ``configuration'' delay of the routing
policies can be minimized.

Also, the duration of a traffic flows plays a critical role. If a
traffic flow is short, i.e., there are only few students in a group,
then SDN cannot recover from the delays suffered for obtaining the
routing policies from the controller. The class can experiment with
varying group sizes, e.g., have only two or four students at each source
node \texttt{A} and \texttt{B} to see if IP will be faster or SDN?
These experiments can be conducted for different distances from node
$a$ to the controller to observe the interactions between the control
plane delay for going to the controller versus the duration of a data
plane traffic flow to make the control plane route optimization pay
off.

Another variation is to make the control plane interactions more
realistic by sending individual runners from the controller to each
intermediate node that is involved in routing a traffic flow.
Specifically, for the \texttt{A} to \texttt{C} traffic flow that means
that after the query from the first encountered intermediate node $a$
has arrived at the controller, and the controller has devised the
optimized routing policy $a$, $b$, and $c$, this routing policy is
sent via three individual ``control plane runners'' from the
controller to the intermediate nodes $a$, $b$, and $c$ so that each of
these intermediate nodes is configured to properly route the
\texttt{A} to \texttt{C} traffic flow.


As an extension, virtualization~\cite{ble2015pai,ble2016con,kil2018pla}
of SDN networks can be considered. The concept of virtualization has rarely been
considered in K-$12$ outreach; specifically, the virtual machine concept
in computing has been considered in~\cite{nas2014usi}. In the present
context of SDN networks, virtualization requires that all control
plane interactions have to traverse a hypervisor. That is a network
node sending a query to the controller first has to traverse a central
hypervisor that may be in a different location than the
controller. Then, from the hypervisor, the query has to proceed to the
controller. On the return, the controller action policy has to
traverse the hypervisor, and then proceed to the node.

\section{English Language Arts Story Writing and Cartooning} \label{elac:sec}
In this section, we outline the design of communication network
outreach activities for integration in the English Language Arts (ELA)
curriculum at middle schools (grades $6$--$8$).  We map the communication
network principles covered in the PE activities in
Section~\ref{peact:sec} into writing and cartooning activities for ELA
classes. The story writing and cartooning activity should first
provide students with a text description of the communication network
principle following the various Background subsections in
Section~\ref{peact:sec} so that students acquire the conceptual and its
functional background on the covered communication network principle.

The middle school students will then be prompted to translate the
dynamics and exchanges in the communication network principle into
cartoon stories.  Data packets of the data traffic flows and control
packets for the SDN control can be represented, i.e., personified, by
cartoon characters. The characters can trace the packets through the
network. For instance, a control packet ``character'' will
traverse the network to the SDN controller to be processed and to
receive instructions (control actions). The character will then carry
the instructions back to the originating SDN switch to execute the
corresponding flow actions, e.g., process data payload traffic
packets. The students should receive feedback on technical and
functional correctness of their cartoon stories from the outreach
team, while the regular English Language Arts teacher should evaluate
the merits of the cartoons with respect to English writing.

Alternatively, the students can be prompted to represent the
communication network dynamics into writing stories.
The stories provide an opportunity for writing assignments that
characterize the personas that represent the various communication
network entities.

\section{Evaluation}  \label{eval:sec}

\subsection{Formative Teacher Feedback}
We interviewed two highly experienced 9th grade teachers,
specifically, one PE and health sciences teacher (21 years of teaching
experience) and one science-biology teacher (40 years of teaching
experience), to obtain formative feedback on the designed outreach
activities. Both teachers are female; the PE teacher teaches co-ed PE
classes with typically 36 students in a class.  The following
subsections summarize the feedback obtain from the teachers.

\subsubsection{Cover Story}
The teachers noted that age-appropriate and timely cover stories are
important to motivate the students. K-12 students currently use text
messaging and the Tiktok Challenge very extensively. The students are
also used to experiencing networking delays when downloading or
uploading videos. Relating the networking principles to these
real-life Internet applications that the students know and value would
help to get the students interested in the engineering
activities. Also, the cover story could involve the outreach activity
facilitators asking the students whether they have ever experienced
delays on the Internet and whether such internet delays have disrupted
their online activities, such as playing multi-player games.

\subsubsection{Workshop Personnel}
The teachers recommended to have the engineering outreach activities
mainly presented and facilitated by engineering students and faculty
that come from a university to the K-12 school. Outside visitors may
draw more of the students' attention. Nevertheless, it is important to
first obtain the buy-in from the individual teachers. Teacher buy-in
is essential for the preparation of the engineering outreach visit and
for monitoring and correcting student behaviors during the outreach
activities.

\subsubsection{Outreach Activity Sequencing}
For the PE activities it is important to keep the introductory part,
i.e., the presentation of the cover story and the explanation of the
underlying engineering principles very short. The students have been
sitting still all day; thus, when they are arriving for the PE class,
they are looking forward to move. The teachers recommended to
structure the outreach activity such that the introductory part is
very short, say three minutes or less. Giving too much detail and
lengthy explanations of the underlying principles before the first
physical activity would likely be counterproductive and demotivate the
students. Ideally, an initial short introductory part should be
immediately followed by physical activities. Then, after the first
physical activity, more detail and explanation of the underlying
concepts can be provided, following my more physical activities, and
then further elaboration of the explanations. Overall, it is important
to scaffold the instructions for the physical activities such that the
students gradually learn to perform more and more of the protocol
operations by themselves.

The teachers also felt that the students may get intimidated by
mathematical formulas. They recommended to present that outreach
activities such that mathematical formulas are avoided. Alternatively,
the formulas can be presented after the completion of a physical
activity to validate the outcomes that the students observed from the
physical activity. If the formulas are used, then it is important to
explain the meaning of each mathematical variable in the corresponding
physical activity.

The teachers also noted that the outreach activities are generally age
appropriate for the range from grades 5 to 9.  However, the issue of
a student being considered as fully arrived after exactly $1/R$ time
has been spent waiting at the node that the student has physically
arrived to as confusing and too complicated for the lower grades
(4--7).  The higher grades (7--10) can likely handle this complexity
and it will make the modeling more realistic and better reflect the
true behavior of a real network for them.

\subsubsection{Competition Aspect}
The teachers emphasized that the students love competition. The
students generally are motivated and try harder if they work towards a
goal in a competitive setting. The teachers recommended to conduct
competitive races in relative quick succession to keep the students
excited and actively moving. Through a quick succession of competitive
races for different parameter settings of the modelled network
architecture and protocol, the students can discover how the different
parameter settings influence the performance of the network.

\subsubsection{Integration of PE and Science Classes}
The teachers suggested that the students could collect measurement
data, e.g., the various delays for different network parameters, in the
PE class. One student from each team can be assigned as record keeper
to measure and record the delay times. The record keeper could be a
student who is injured or cannot participate in physical
activities. The students are generally used to keep track of scores
and times with stop watches and clipboards. Electronics, e.g., smart
phones, are generally not permitted in PE classes as they may break.

Then, in a successive science class, the students could be introduced
to the underlying concepts, theory, and mathematical formulas and
could evaluate the network delays from the formulas. The calculated
delays could then be compared with the delays measured in the PE
class.

\subsection{Perception Survey}
This section presents a survey to evaluate the student perceptions
interest, utility (importance), self-efficacy, and negative
stereotypes related to network engineering.  The presented evaluation
survey has been adapted from an extensive evaluation of the
engineering outreach activities for middle school students in the
Arizona Science Lab~\cite{inn2012ari,ozo2019sch}. The construct
validity and internal reliability of the survey had been assessed
following~\cite{aik2005psy}.

The goal of the
evaluation is to provide a critical understanding of the perceptions
that middle school students have about the introduced network
engineering outreach activities.  A thorough understanding of these
perceptions is a prerequisite for the formative and summative
evaluation of network engineering outreach activities that are
effective in the sense of enhancing the likelihood that middle school
students will pursue programs of study and careers in network
engineering.

The adapted evaluation survey consists of four constructs, each with
three items:
\begin{enumerate}
\item Interest in Network Engineering
\begin{enumerate}
\item I would like to learn more about network engineering
\item	I would be interested in working as an network engineer
\item I would be interesting in studying network engineering at a university
\end{enumerate}
\item Self-efficacy in Network Engineering
\begin{enumerate}
\item I could succeed in network engineering
\item	I believe I have talent for network engineering
\item	I would get good grades in network engineering classes
\end{enumerate}
\item Negative Stereotypes
\begin{enumerate}
\item Only nerds spend a lot of time doing network engineering.
\item	Network engineers are unpopular people
\item	Network engineers are boring people
\end{enumerate}
\item Importance of Network Engineering:
\begin{enumerate}
\item Network engineering plays an important role in solving society's problems.
\item	Network engineers make people's lives better.
\item Network engineering affects our everyday lives.
\end{enumerate}
\end{enumerate}
In addition, the pre-survey will collect demographic information,
namely age, gender, and ethnicity.
The evaluation survey can be utilized to evaluate the student
perception prior to the network engineering outreach activity, i.e.,
as a pre-survey.
In addition, the evaluation survey can be utilized to
evaluate the student perceptions as a post-survey
right after the conclusion
of the network engineering outreach activities, and if logistics
permit, after a few weeks as a delayed-post-survey.

\section{Conclusion}
We have designed a K-$12$ engineering outreach program covering the
content area of communication network engineering, specifically, the
elementary principles of packet switched networks. The outreach
program is based on physical education (PE) activities as well as
English Language Arts (ELA) assignments.  We have collected formative
evaluation feedback from network engineering and K-$12$ instruction
experts to refine the outreach program an make it suitable for the
various age groups, e.g., upper elementary school grades ($4$ and $5$)
through the middle of middle school grades ($6$ and $7$) as well as the
middle school grades ($7$ and $8$) through the lower high school grades ($9$
and $10$). A summative evaluation survey of the impact of the outreach
program on student interest, self-efficacy, utility, and negative
stereotype perceptions has been adapted from prior engineering outreach
research.

There are several important directions for future work.  The developed
outreach program should be piloted with the various age groups (e.g.,
grades $4$--$7$ and grades $7$--$10$) and further refined and adapted to these
age groups. Then extensive evaluations with diverse students
populations of these two age groups should be conducted with the
provided student perception survey to evaluate pre-program to
post-program changes in the student perceptions.

Another future work direction is to expand the covered communication
network principles. The activities designed so far have focused on
elementary principles. Future work can expand the covered principles
to techniques that enhance the communication network performance or
make communication networks more secure, as network security is a
topic that most K-$12$ students find highly important. For instance,
future work can expand the covered principles to network
coding~\cite{bas2013net,far2014sur,gab2018cat,luc2018ful,nae2017net,ngu2020dse,wun2019pro},
which can enhance performance as well as enhance security.
The coding at the source, could for instance be represented by wrapping
a bucket that has some holes with fabric to slow the outflow of water
and thus the loss of water during the transport over the network.
The more time that the students spent at the source wrapping the
leaky bucket in fabric, the less water will be lost during transport.
The
principle of coding has so far mainly been covered in hacking
workshops, see e.g.,~\cite{cha2019sec,kon2018exp}, in the K-$12$ age
group. Another direction that is appealing as many students are
interested in the latest advances in smart-phones and similar mobile
computing nodes, is to incorporate the principle of accelerators for
specific networking related functions~\cite{lin2019sur,nie2017fpg,sha2020har}.

\section*{Acknowledgment}

\bibliographystyle{IEEEtran}


\begin{IEEEbiography}[{\includegraphics[width=1in,height=1.25in,clip,keepaspectratio]{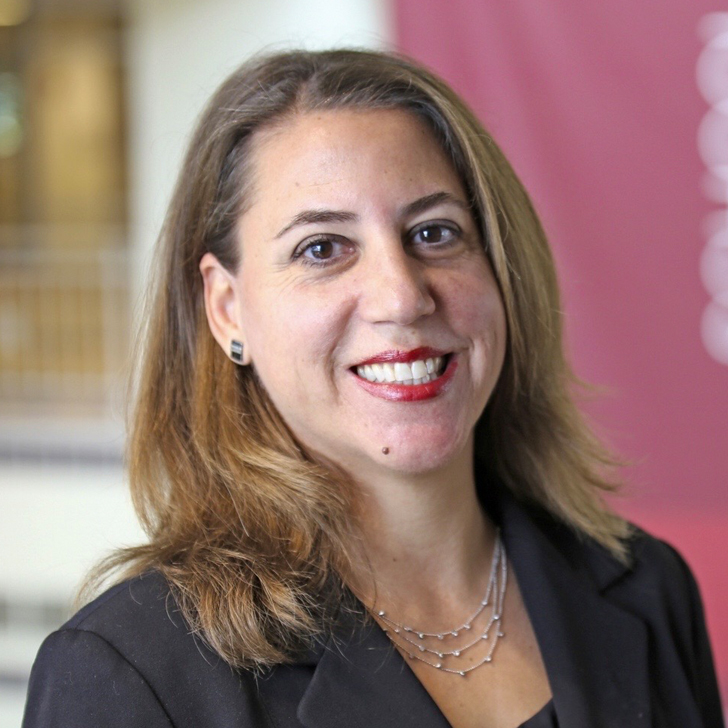}}]{Gamze Ozogul}
is an Associate Professor with the Instructional Systems and
Technology Department in the School of Education at Indiana
University--Bloomington.  From 2013 to 2019, she was an Assistant
Professor at Indiana University. She received the Ph.D. in Educational
Technology from the Division of Psychology of Education, Arizona State
University, Tempe, in 2006, the M.S. in Computer Education and
Instructional Technology from Middle East Technical University,
Ankara, Turkey, in 2002, and the B.S. in Curriculum and Instruction
from Hacettepe University, Ankara, Turkey, in 2000.

\end{IEEEbiography}

\begin{IEEEbiography}[{\includegraphics[width=1in,height=1.25in,clip,keepaspectratio]{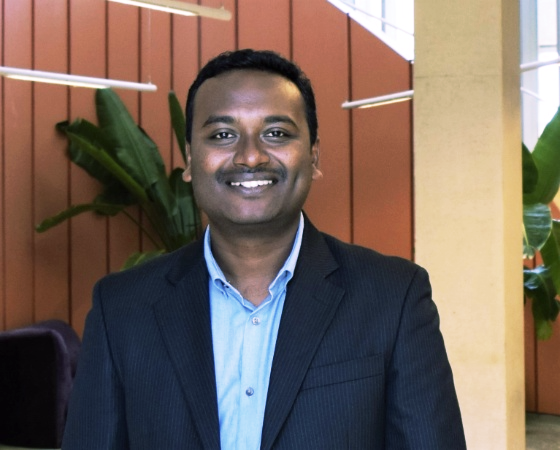}}]{Akhilesh S. Thyagaturu}
is an Sr. Software Engineer at Programmable Solutions Group, Chandler,
AZ, USA, and an Adjunct Faculty in the School of Electrical, Computer,
and Energy Engineering at Arizona State University (ASU), Tempe. He
received the Ph.D. in Electrical Engineering from Arizona State
University, Tempe, in 2017.  He serves as reviewer for various
journals, including the \textit{IEEE Communications Surveys \&
  Tutorials}, \textit{IEEE Transactions of Network and Service
  Management}, and \textit{Optical Fiber Technology}. He was with
Qualcomm Technologies Inc., San Diego, CA, USA, as an Engineer from
2013 to 2015.
\end{IEEEbiography}

\begin{IEEEbiography}[{\includegraphics[width=1in,height=1.25in,clip,keepaspectratio]{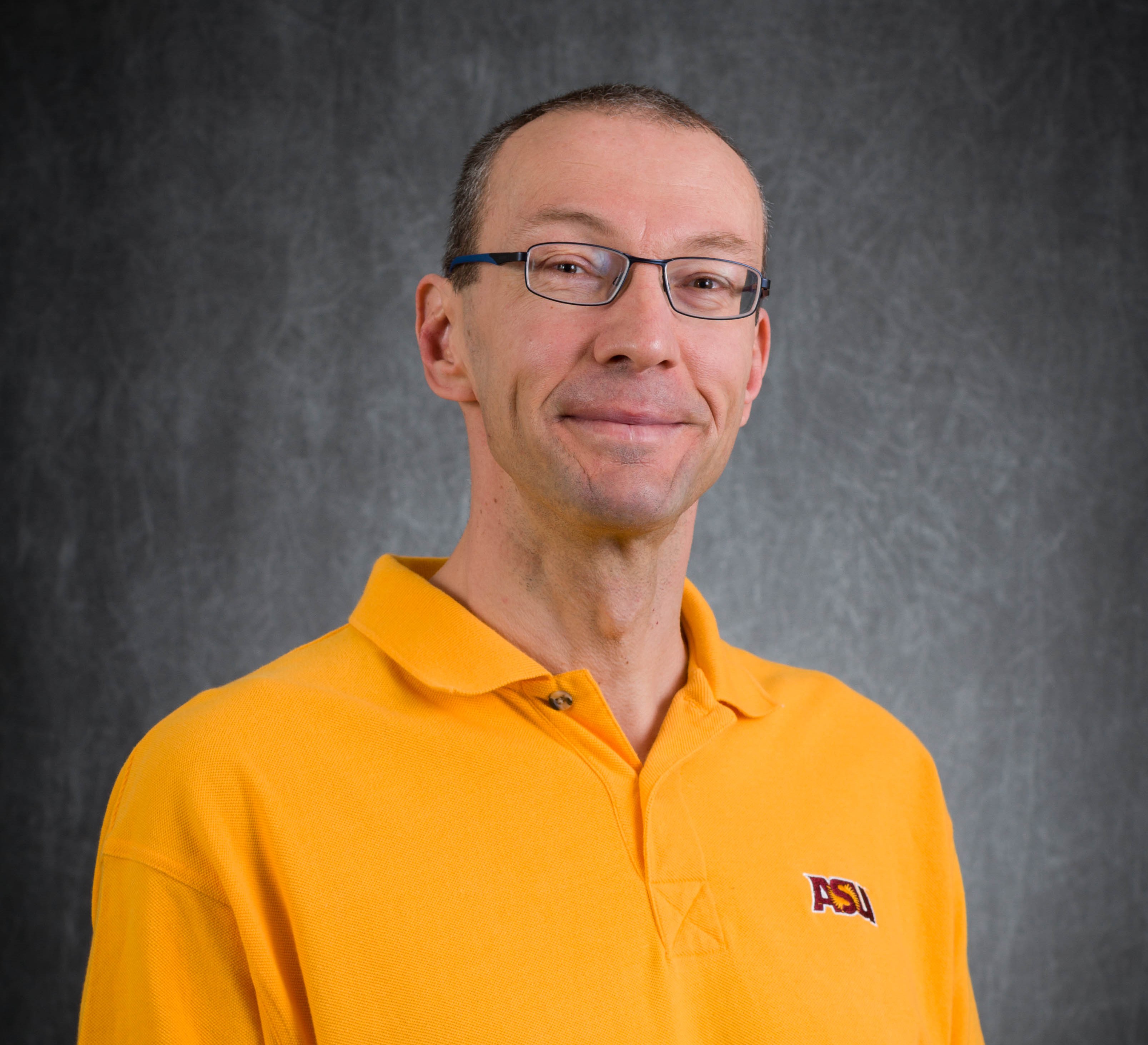}}]
{\textbf{Martin Reisslein}}~(S'96-M'98-SM'03-F'14) is a Professor in
the School of Electrical, Computer, and Energy Engineering at Arizona
State University (ASU), Tempe.  He received the Ph.D. in systems
engineering from the University of Pennsylvania, Philadelphia, in
1998. He currently serves as Associate Editor for the \textit{IEEE
  Transactions on Mobile Computing}, the \textit{IEEE Transactions on
  Education}, and \textit{IEEE Access}, as well as \textit{Computer
  Networks}. He is an Associate Editor-in-Chief of the \textit{IEEE
  Communications Surveys \& Tutorials} and a Co-Editor-in-Chief of
\textit{Optical Switching and Networking}. He chaired the steering
committee of the \textit{IEEE Transactions on Multimedia} from
2017--2019 and was an Associate Editor of the \textit{IEEE/ACM
  Transaction on Networking} from 2009--2013. He received the IEEE
Communications Society Best Tutorial Paper Award in 2008, a Friedrich
Wilhelm Bessel Research Award from the Alexander von Humboldt
Foundation in 2015, as well as a DRESDEN Senior Fellowship in 2016 and
in 2019.
\end{IEEEbiography}

\begin{IEEEbiography}[{\includegraphics[width=1in,height=1.25in,clip,keepaspectratio]{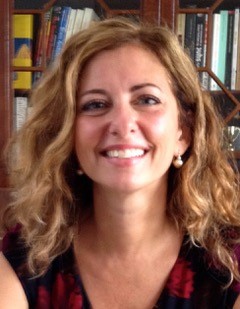}}]
 {Anna Scaglione} (F'11)~(M.Sc.'95, Ph.D. '99) is professor in
 Electrical and Computer Engineering at Arizona State University,
 Tempe. She was Professor of Electrical Engineering previously at UC
 Davis (2010--2014), Associate Professor at UC Davis 2008--2010 and at
 Cornell University (2006--2008), and Assistant Professor at Cornell University (2001--2006)
 and at the University of New Mexico (2000--2001). Dr. Scaglione's
 research is in modeling, analyzing, and designing mechanisms for
 reliable and efficient networked systems. Her expertise is in
 statistical signal processing, wireless communications, and energy
 delivery systems. She is the recipient of the 2000 IEEE Signal
 Processing Transactions Best Paper Award, the 2013, IEEE Donald
 G. Fink Prize Paper Award, and, with her student, of the 2013 IEEE
 Signal Processing Society Young Author Best Paper Award.
\end{IEEEbiography}

\end{document}